# Tailoring Thermal Radiative Properties with Film-Coupled Concave Grating Metamaterials

Hao Wang and Liping Wang[*]

School for Engineering of Matter, Transport & Energy
Arizona State University, Tempe, Arizona, 85287 USA

*Corresponding author: liping.wang@asu.edu

**ABSTRACT**

This work numerically investigates the radiative properties of film-coupled metamaterials made of a two-dimensional metallic concave grating on a continuous metal film separated by an ultrathin dielectric spacer. Spectrally-selective absorption is demonstrated in the visible and near-infrared regime, and underlying mechanisms are elucidated to be either localized magnetic polaritons (MPs) or surface plasmon polaritons (SPPs). The unique behaviors of MPs and SPPs are explained with the help of electromagnetic field distributions at respective resonance frequencies. An inductor-capacitor model is utilized to further confirm the excitation of MP, while dispersion relation is used to understand the behaviors of different SPP modes. Geometric effects of ridge width and grating period on the resonance absorption peaks are discussed. Moreover, directional responses at oblique incidences for different polarization states are studied. Fundamental understanding gained here will facilitate the design of novel metamaterials in energy harvesting and sensing applications.

# 1. Introduction

Spectral control of thermal radiation has numerous applications in energy harvesting [1-3], thermal management [4], radiative cooling [5], and optical sensing [6, 7]. Metamaterials, which refer to a class of artificial materials with exotic electromagnetic properties, have been studied for tailoring thermal radiative properties [8]. Metamaterials are usually made of micro/nanostructures with feature sizes smaller than the wavelength of incident light. Recently, film-coupled metamaterials, consisting of subwavelength periodic patterns on a dielectric layer and a metal ground plane, have drawn much attention [9]. Selective metamaterial absorbers or emitters with different patterns such as one-dimensional (1D) grating [10, 11], 2D disk arrays [12], patch arrays [13-15], cross bars [16], and nanoparticles [17-20] have been investigated. Excitations of magnetic polaritons (MPs) [21-23] and surface plasmon polaritons (SPPs) [24, 25] have been demonstrated as the main mechanisms responsible for the spectral selectivity in thermal absorption or emission in these metamaterials. MP refers to the coupling between external electromagnetic fields and magnetic resonance insides the structures, while SPP stems from collective oscillation of charges at the interface of dissimilar materials whose real parts of permittivity have opposite signs [26].

On the other hand, radiative properties of film-coupled concave grating structures with 2D periodic mesh-like patterns are little studied. Gou et al. [27] investigated spectral optical properties of a metal-insulator-metal stack with rectangular hole arrays but mainly focused on coupled SPP modes across different layers. Lee et al. [28] introduced a concave grating structure made of nickel, in which selective absorptance is explained by excitation of cavity resonance. There have been discussions on photonic crystals [29] and fishnet structures [30, 31] with similar geometries, but the effects of MP and SPP in controlling the radiative properties of film-coupled concave gratings are not yet well understood or systemically studied by far.

In this work, we numerically investigate effects of magnetic and surface plasmon polaritons in tailoring the radiative properties of a film-coupled concave grating metamaterial structure with finite-difference time-domain (FDTD) simulation. Selective absorption in the visible and near-infrared regime is demonstrated, and electromagnetic fields are plotted to elucidate underlying mechanisms responsible for each absorption peak. Geometric effects of ridge width and grating period are also discussed. An inductor-capacitor (LC) circuit model and the SPP dispersion relation are utilized to theoretically predict MP and SPP resonance conditions. Moreover, optical responses at oblique incidences are investigated for both transverse magnetic (TM) and transverse electric (TE) polarized waves to gain a full understanding in MP and SPP behaviors.

## 2. Numerical Method

Figure 1(a) depicts the periodic film-coupled metamaterial structure under investigation, which is made of an aluminum (Al) concave grating on a silicon dioxide (SiO$_2$) spacer and Al substrate. Note that Al is chosen as the metal material due to its natural abundance. The 2D periodic mesh-like grating, which is considered to be symmetric in x and y directions for simplicity, has a cavity width *b*, ridge width *w*, grating period Λ, and grating height *h*, while the SiO$_2$ spacer thickness is *t*. An electromagnetic wave with a wavevector k$_{inc}$ is incident onto the structure at an incidence angle *θ*, which is defined as the angle between k$_{inc}$ and surface normal. The plane of incidence (POI) is spanned by incident electric field vector **E** and k$_{inc}$, and is taken to be parallel to the x-z plane (i.e., y-direction incident wavevector $k_{y,\text{inc}} = 0$), for which the conical diffraction would not occur. The electric field vector **E** is in the POI with TM polarized waves, as shown in the figure, but is perpendicular to the POI for TE waves. Note that the x-direction incident wavevector $k_{x,\text{inc}} = k_{\text{inc}} \sin\theta = 0$ at normal incidence. A set of base geometric values, such as Λ = 1 μm, *b* = 0.75 μm, *w* = 0.25 μm, and *h* = *t* = 30 nm, was considered for the film-coupled concave grating structure.

The FDTD simulation was implemented using a commercial package (Lumerical Solutions, Inc.). Frequency-dependent optical constants of both Al and SiO$_2$ were obtained from Palik's tabular data [32]. Manual refined meshes with sizes of 5 nm in both *x* and *y* direction, and 1 nm in *z* direction were used to ensure the numerical convergence. Periodic boundary condition was applied in both *x* and *y* directions at normal incidence to reduce the 3D simulation domain into one unit cell, while Bloch boundary condition had to be used at oblique incidences to account for the phase difference in neighboring unit cells. Perfect matched layers with a reflection coefficient less than $10^{-6}$ were placed along z direction. A linearly polarized plane wave source was positioned at 1.2 μm above the structure surface, while spectral reflectance $R$ was obtained by a frequency-domain power monitor placed at 1.5 μm above the structure. Note that the 200-nm-thick Al substrate is optically opaque. Therefore, the spectral-directional absorptance can be readily calculated by $\alpha = 1-R$ according to the energy balance. The frequency range of interest is from 5000 $cm^{-1}$ to 25000 $cm^{-1}$ in wavenumber with a spectral resolution of 10 $cm^{-1}$.

## 3. Results and Discussion

3.1 Radiative Properties at Normal Incidence

The spectral normal absorptance of the film-coupled concave grating structure with the set of base geometric values is presented in Fig. 1(b) at TM-wave incidence, though TE waves will yield the same results at normal incidence due to the geometric symmetry. Absorption peaks are observed at the several frequencies of 8350 $cm^{-1}$ (with absorptance $\alpha$ = 0.89), 9970 $cm^{-1}$ ($\alpha$ = 0.80), 13740 $cm^{-1}$ ($\alpha$ = 0.36), 18870 $cm^{-1}$ ($\alpha$ = 0.43), 21150 $cm^{-1}$ ($\alpha$ = 0.23), and 23580 $cm^{-1}$ ($\alpha$ = 0.36). These peaks were respectively labeled as MP1, SPP(±1,0), SPP(±1,±1), SPP(±2,0), SPP(±2,±1), and MP3 from lower to higher frequencies, which are associated with the excitations of different MP and SPP modes. The underlying mechanisms and unique behaviors of SPPs and MPs are to be discussed in following sections. Table 1 summarizes the resonance wavenumbers for different MP and SPP modes obtained from FDTD simulation. Note that in Fig. 1(b), the small absorption peak around 5500 cm$^{-1}$ is due to the SPP excited at the SiO$_2$-Al interface, which will not be discussed in detail. The present study will focus on the SPP modes at higher frequencies, which are excited at the air-Al interface.

The spectral normal absorptance for the concave grating structure without the SiO$_2$ spacer but with the same dimensions is also plotted for comparison. It is found that without the dielectric spacer, MP1 and MP3 absorption peaks disappear, while all the SPP peaks still exist at nearby resonance frequencies, indicating that a dielectric spacer is essential for excitation of MPs but not for SPPs in film-coupled concave grating metamaterials. This is the same behavior that has been previously discussed on the excitation of MPs in film-coupled 1D gratings [10, 21] and 2D convex gratings [14, 15]. The dielectric spacer functions as a capacitor that forms a resonant LC circuit to excite the magnetic resonance. The top and bottom metals would be short circuited without the dielectric spacer, which disables the excitation of magnetic resonance at a particular frequency.

To help understand the excitation of MPs and SPPs in film-coupled concave gratings, the spectral absorptance of a film-coupled 1D grating with x-direction periodicity (i.e., 1D grooves/ridges along y direction) and the same geometric parameters is calculated and shown in Fig. 1(b). It can be observed that both MP1 and MP3 absorption peaks appear at almost the same frequencies for film-coupled 2D concave grating and its 1D counterpart, suggesting that the excitation of MPs inside the film-coupled concave grating is similar to 1D grating at these frequencies. On the other hand, when the structure changes from 1D to 2D, spectral peaks due to SPP($\pm1,0$) and SPP($\pm2,0$) stays, but additional SPP modes, i.e., SPP($\pm1,\pm1$) and SPP($\pm2,\pm1$), emerge as well, indicating the effect of grating periodicity in the y direction on exciting SPP modes.

3.2 Electromagnetic Field Distribution at Excitation of Localized MPs

To explain underlying mechanism for these absorption peaks associated with MPs, the distribution of electromagnetic fields at different sections inside the structure are illustrated in Fig. 2, with arrows symbolizing electric field vectors and contour representing the magnetic field strength as the logarithm of the square of magnetic field normalized to the incidence (i.e., $\log_{10}|H/H_{inc}|^2$). Figure 2(a) indicates the x-y cross section at the center plane of the dielectric spacer (i.e., $z = 0$) where the electromagnetic fields were plotted for MP1 in Fig. 2(c) and MP3 in Fig. 2(e), while Fig. 2(b) depicts the x-z cross section at the location of half period (i.e, $y = \Lambda/2$) where the fields were shown for MP1 in Fig. 2(d) and MP3 in Fig. 2(f).

At the MP1 resonance, magnetic field is greatly enhanced by more than two orders of magnitude under the y-direction ridges only in the section between left and right cavities (i.e., discontinuous region along x direction), but is suppressed under the x-direction ridges (i.e., continuous region), as shown in Fig. 2(c). As further observed in Fig. 2(d), the magnetic field enhancement occurs inside the dielectric spacer only underneath top metallic ridges, accompany by an induced electric current loop. The electromagnetic field distribution shows the exact behavior of magnetic resonance or MP as previously studied in film-coupled 1D grating structures [10, 14, 15]. The internal magnetic resonance is excited by the external electromagnetic waves at a particular frequency, and the coupling leads to the collective oscillation of charges, which forms a current loop and induces a resonant magnetic field, resulting in strong energy absorption inside the structure.

The excitation of magnetic resonance inside the film-coupled concave gratings can be easily understood with the help of its 1D counterpart, in which the behaviors of MPs have been intensively discussed [10, 14, 15]. After all, the 2D concave grating can be treated as superposition of two 1D gratings in x and y directions, and thus MPs would occur at similar frequencies, as confirmed by the numerical simulation results in Fig 1(b). However, the magnetic resonance only occurs at the discontinuous region between cavities in the concave grating, but exists continuously along the entire ridges in film-coupled 1D gratings. This can be explained by the essential role of cavities, or grooves in 1D gratings, which separate the metals and function as capacitors in enabling excitation of magnetic resonance. At the continuous region without the cavities, magnetic resonance cannot be excited under the continuous x-direction ridges, as the oscillating charges cannot accumulate to form capacitors and inductors, which thereby fails to excite magnetic resonance. The reason why magnetic resonance is excited under y-direction ridges rather than x-direction ones is because the magnetic field is along y-direction with TM polarized waves considered in this case. Oscillating electric current that induces magnetic resonance has to be excited inside the plane that is perpendicular to the incident magnetic field. Due to the 2D geometry, magnetic resonance and the resulting absorption peaks are expected to occur for TE polarized waves.

Figures 2(e) and 2(f) show the electromagnetic field distribution at MP3 resonance, which is the third harmonic mode of MP featured with three antinodes in localized magnetic fields and three electric current loops with alternating directions. Similarly, the magnetic resonance with MP3 occurs only under y-direction ridges at the section between cavities, and is disabled under the continuous x-direction ridges. Note that the second harmonic mode MP2 is not observed here because the magnetic field strengths of two antinodes with opposite directions cancel each other due to the symmetry at normal incidence [11].

3.3 An LC Model for Predicting Resonance Conditions of MPs

LC circuits, which stem from the charge distributions at magnetic resonance, have been employed to successfully confirm the excitation of magnetic resonance in film-coupled 1D gratings [10]. In fact, the LC model (per unit length along y direction), as sketched in Fig. 3(a), can be directly used for theoretically predicting resonance condition of MPs the film-coupled concave grating metamaterials. For the sake of completeness, the relevant capacitors and inductors will be described again for film-coupled concave gratings. $C_\mathrm{m}$ is the parallel-plate capacitor between top grating ridge and metal substrate, $C_\mathrm{cavity}$ is the cavity capacitor between neighboring ridges, $L_\mathrm{m}$ and $L_\mathrm{k}$ are respectively the parallel-plate inductor and the kinetic inductor. The expressions for $C_\mathrm{m}$, $C_\mathrm{cavity}$, $L_\mathrm{m}$ and $L_\mathrm{k}$ are listed below:

$$C_\mathrm{m} = \frac{c_1 \varepsilon_0 \varepsilon_\mathrm{d} w}{t} \tag{1}$$

$$C_\mathrm{cavity} = \frac{\varepsilon_0 h}{b} \tag{2}$$

$$L_\mathrm{m} = 0.5 \mu_0 w t \tag{3}$$

$$L_\mathrm{k} = \frac{-w}{\omega^2 \varepsilon_0 \varepsilon'_\mathrm{m} \delta} \tag{4}$$

where $\omega$ is the angular frequency, $c_1$ is a coefficient accounting for non-uniform charge distribution at the metal surface and is taken as 0.2 in the calculation, $\varepsilon_0$ is the permittivity of vacuum, $\varepsilon_\mathrm{d}$ is the permittivity of the SiO$_2$ spacer, $\mu_0$ is the permeability of vacuum, $\varepsilon'_\mathrm{m}$ is the real part of the permittivity of the metal, and $\delta$ is the penetration depth of the metal.

The in-series LC circuit in Fig. 3(a) can be further simplified into a closed form shown in Fig. 3(b) due to grating periodicity. The total impedance of the LC circuit is:

$$Z_{total} = \frac{L_m + L_k}{1 - \omega^2 C_{cavity}(L_m + L_k)} - \frac{2}{\omega^2 C_m} + (L_m + L_k) \qquad (5)$$

Magnetic resonance occurs when the circuit has zero impedance, and therefore, the resonance condition of MP1 can be solved by setting $Z_{total} = 0$. With the base geometric values for the film-coupled concave grating, the LC model predicts the MP1 to occur at the frequency of 7,852 cm$^{-1}$, which matches well the FDTD simulation with a relative difference of 5.3%, confirming the excitation of magnetic resonance inside film-coupled concave grating metamaterials. The resonance frequency of MP3 can be simply approximated to be three times of MP1 resonance frequency as $\omega_{MP3} = 3\omega_{MP1}$ when $C_{cavity} \ll C_m$, which is valid in this case. Comparison on the resonance wavenumbers for MP1 and MP3 between the LC model and the FDTD simulation is summarized in Table 1.

3.4 Electromagnetic Field Distribution and the Dispersion Relation of SPPs

To understand the underlying mechanism for the absorption peaks related to the SPP modes in film-coupled concave gratings, electromagnetic field distribution was plotted at SPP($\pm$1,0) in the prescribed x-y and x-z cross sections in Figs. 4(a) and 4(b), respectively. The magnetic field shows maximum and minimum values in a periodic pattern along x direction, while is uniform in the y direction. Besides, the electric field vectors display a periodic oscillating pattern in the z direction at the air-grating interface. In addition, the magnetic field also decays quickly away from the interfaces into the metal. The field distribution shows the exact behaviors of a surface wave along the interface when a SPP mode is excited. In this case, the surface waves propagate only in the x direction with SPP($\pm$1,0). At the SPP($\pm$1,$\pm$1) mode, surface waves are not only excited in the x direction but also in the y direction, resulting in a symmetric and periodic field pattern in the 2D plane as shown in Figs. 4(c) and 4(d).

SPP can be excited at an interface between two dissimilar materials with permittivities $\varepsilon_1$ and $\varepsilon_2$ when its in-plane momentum or wavevector $k_{SPP}$ satisfies the dispersion relation expressed as [26]:

$$|k_{SPP}| = \frac{\omega}{c_0}\sqrt{\frac{\varepsilon_1 \varepsilon_2}{\varepsilon_1 + \varepsilon_2}} \qquad (6)$$

SPP dispersion curve usually lies at the right side of light line, and thereby can only be excited by evanescent waves at a planar interface. However, SPP can be excited in periodic gratings because large in-plane wavevectors can be matched for diffracted waves according to the grating Bloch condition:

$$k_{\parallel,mn} = \left(k_{x,\text{inc}} + \frac{2\pi m}{\Lambda_x}\right)\hat{x} + \left(k_{y,\text{inc}} + \frac{2\pi n}{\Lambda_y}\right)\hat{y} \qquad (7)$$

where m and n are diffraction orders respectively in x and y directions, while $\Lambda_x$ and $\Lambda_y$ represent grating period in x and y directions (i.e., $\Lambda_x = \Lambda_y = \Lambda$ in the present study). The resonance frequencies for different SPP modes can be theoretically obtained by solving $\left|k_{\mathrm{SPP}}\right|=\left|k_{\|,mn}\right|$ with different diffraction orders (*m,n*) in x and y directions.

Comparison is made on the resonance frequencies of different SPP modes at the air-Al interface in the film-coupled concave gratings under normal incidence (i.e., $k_{x,\mathrm{inc}}=k_{y,\mathrm{inc}}=0$) between the FDTD simulation and the SPP dispersion relation prediction, as summarized in Table 1. It can be seen that dispersion relation makes accurate predictions with relative error less than 5% for four SPP modes observed in Fig. 1(b), namely SPP(±1,0), SPP(±1,±1), SPP(±2,0), and SPP(±2,±1). Note that SPP(±m,0) modes can only be excited for TM incidence. This is because at excitation of SPP(±m,0), 2D gratings can be equivalent to 1D gratings with x-direction periodicity, which only contributes to the in-plane wavevector $k_{\|,mn}$. In such a 1D grating with periodicity in x direction, SPP can be only excited with TM incidence. Similarly, SPP(0,±n) can be excited only for TE waves. Other SPP(±m,±n) modes can be excited in the concave grating for either TE or TM incidence, as both x- and y-direction grating periodicity could make contribution to the in-plane wavevector. Similar SPP behaviors have been also discussed in 2D convex gratings [14]. In fact, the discussion can be applied to any types of 2D grating structures with the same grating periods, no matter whether it is convex, concave, or other geometric patterns, as only the 2D grating periodicity affects the SPP behaviors.

3.5 Geometric Dependences of MP and SPP Resonances

In order to understand the effect of geometric parameters on the SPP and MP behaviors as well as the absorption peaks, spectral normal absorptance of the film-coupled concave grating was calculated with independently varying grating ridge width $w$ and period $\Lambda$, as respectively presented in Figs. 5(a) and 5(b), while all other parameters remain as the base values. When the ridge width $w$ changes from 0.2 μm to 0.3 μm, absorption peak due to MP1 shifts to lower frequencies, which can be understood by the increase of capacitance and inductance values according to Eqs. (1-4). This results in a smaller magnetic resonance frequencies as described by the LC circuit model. On the other hand, the SPP($\pm1,0$) and SPP($\pm1,\pm1$) peaks stay at the same spectral locations, as SPP resonance conditions only depend on the grating period, rather than ridge width, as discussed with the dispersion relation. Similar with MP1, the MP3 peak also shifts to lower frequencies with larger ridge width, and influences nearby SPP($\pm2,0$) and SPP($\pm2,\pm1$) modes, both of which in turn slightly shift to lower frequencies due to the interaction with MP3.

Figure 5(b) shows the effect of grating period with values of 0.9 μm, 1 μm, and 1.1 μm on the spectral normal absorptance. As expected, with larger grating period, all the SPP absorption peaks shift to lower frequencies due to smaller $k_{\parallel,mn}$ values suggested by Eq. (7). On the other hand, the MP1 absorption peak frequency barely changes. Note that, the grating period only affects the cavity capacitance $C_{\text{cavity}}$ via cavity width $b = \Lambda - w$. However, with considered geometric values, $C_{\text{cavity}}$ is two orders of magnitude smaller than $C_{\text{m}}$, which is dominant by the ridge width. Therefore, the grating period has little effect on the MP1 resonance condition. In addition, the MP3 absorption peak exhibits a noticeable shift in terms of resonance frequency, which results from the interaction with the nearby SPP($\pm 2, \pm 1$) mode.

In order to further illustrate the geometric effects on the MP and SPP modes, the resonance conditions are compared between the FDTD simulation and LC model prediction for MP1 or dispersion relation for SPP($\pm 1, 0$) as a function of ridge width or grating period in Figs. 6(a) and 6(b), respectively. Clearly, as predicted by the LC model, the MP1 resonance frequency monotonically decreases with wider grating ridges but is almost independent on the grating period. On the other hand, SPP dispersion relation predicts the opposite behaviors for SPP($\pm 1, 0$), whose resonance condition does not change with ridge width but decreases with larger grating period. Both LC model and SPP dispersion predictions on the MP1 and SPP($\pm 1, 0$) behaviors show reasonable agreement with the numerical results from FDTD simulations, which undoubtedly confirms the fact that the resonance absorption peaks in film-coupled concave grating metamaterials are attributed to the excitations of multiple MP and SPP modes. Note that, the differences between theoretical prediction and numerical simulation are mainly caused by the interaction between different MP and SPP modes, which is captured by the simple theoretical models. However, the LC model and dispersion relation are of great benefits in understanding the physical mechanisms and behaviors of MPs and SPPs, and provide useful guidance in designing novel metamaterials for practical applications.

3.6 Behaviors of MPs and SPPs at Oblique Incidences

In order to gain a understanding in the behaviors of MPs and SPPs for controlling radiative properties at oblique incidences, contour plots of the spectral absorptance of the film-coupled concave grating as a function of wavenumber and $x$-component wavevector $k_x$ were calculated for TM and TE polarized waves, as shown in Figs. 7(a) and 7(b) respectively. Considering the possible issue in the numerical convergence at large oblique incidence in the FDTD simulation, the incidence angle $\theta$ is thus limited from $0^o$ to $30^o$. Note that the results at $k_x = 0$ is for normal incidence in the contour plots.

As shown for TM waves in Fig. 7(a), a flat resonance band at MP1 resonance frequency was observed, indicating that the magnetic resonance condition does not change at oblique incidence angles. The diffuse-like behavior of magnetic resonance or MPs has been discussed in film-coupled 1D grating structures before [14, 21], which can be understood by the fact that incidence magnetic field is always along y direction no matter what incidence angle is. Note that the MP1 resonance band is split into two modes by the SPP(−1,0) mode, while MP3 at higher frequencies is divided into three modes due to the crossing between multiple higher-order SPP modes. However, for TE waves shown in Fig. 7(b), both MP1 and MP3 exhibit similar behaviors with flat resonance bands but are continuous without splitting because of different SPP behaviors at TE waves, which do not cause any intersection with MP modes.

There exist more SPP modes at oblique angles other than four at normal incidence for TM waves as shown in Fig. 7(a). In fact, all the SPP modes at normal incidence split into two modes with a high-frequency branch and a low-frequency branch with larger $k_x$. The inclined SPP curves show strong selectivity and dependence in both frequency and $k_x$ or oblique incidence angle. The splitting behavior of SPP modes at oblique incidence can be explained by the SPP dispersion relation coupled with the diffractive nature of periodic gratings, i.e, $|k_{\text{SPP}}| = |k_{\parallel,mn}|$, described in Eqs. (6) and (7). Since $k_{y,\text{inc}} = 0$ in the present study, the in-plane wavevector becomes $|k_{\parallel,mn}| = \sqrt{(k_x + 2m\pi/\Lambda)^2 + (2n\pi/\Lambda)^2}$. Therefore, at oblique incidence with a nonzero $k_x$, either +$m$ or −$m$ for given $m$ and $n$ values will yield increased or decreased $|k_{\parallel,mn}|$ values, leading to two solutions or SPP branches, which merge into one mode at normal incidence. For TE waves shown in Fig. 7(b), the SPP mode does not split into two branches at $k_x \neq 0$ or oblique angles, because either +$n$ or −$n$ will result in the same $|k_{\parallel,mn}|$ value due to the prescribed condition $k_{y,\text{inc}} = 0$. Finally, all the SPP modes have been marked in both contour plots to clearly show their behaviors at oblique incidences for different polarizations.

**4. Conclusion**

In sum, we have numerically demonstrated that thermal radiative properties can be tailored with film-coupled concave grating metamaterials by excitation of multiple MP and SPP resonance modes, which result in selective absorption at the visible and near-infrared frequencies. The physical mechanisms of MPs and SPPs have been elucidated with electromagnetic field distribution. The underlying physics responsible for the selective absorption has been further confirmed with the LC model and SPP dispersion relation. It is found that, MP resonance condition can be tuned by ridge widths, while the SPP peak frequencies can be controlled with the grating period. Besides, MPs show direction-independent behavior, while SPP modes exhibit strong angular selectivity in resonance conditions. Different from film-coupled 1D gratings, both MP and SPP resonance modes can be excited for both TM and TE polarized waves. The fundamental understanding gained from this work will facilitate the novel design of film-coupled metamaterials for energy and sensing applications.

**Acknowledgement**

We would like to thank the supports from the *US-Australia Solar Energy Collaboration - Micro Urban Solar Integrated Concentrators (MUSIC)* project sponsored by the *Australian Renewable Energy Agency (ARENA)*, ASU New Faculty Startup fund, and the seeding fund by ASU Fulton Schools of Engineering.

Table 1. Comparison on the resonance frequencies of different MP and SPP modes at normal incidence for the film-coupled 2D concave grating metamaterial between FDTD simulation and the LC model for MPs or dispersion relation for SPPs.

| Resonance Wavenumber (cm$^{-1}$) | MP1 | MP3 | SPP ($\pm1$,0) | SPP ($\pm1$,$\pm1$) | SPP ($\pm2$,0) | SPP ($\pm2$,1) |
|---|---|---|---|---|---|---|
| FDTD Simulation | 8,290 | 23,580 | 9,970 | 13,740 | 18,870 | 21,150 |
| LC model for MP | 7,852 | 23,556 | - | - | - | - |
| SPP Dispersion Relation | - | - | 9,949 | 14,070 | 19,898 | 22,247 |
| Relative Error (%) | 5.3% | 0.1% | 0.2% | 2.4% | 5.5% | 5.2% |

**Figure Captions:**

Fig. 1   (a) Schematic of a periodic film-coupled concave grating metamaterial considered in the present study. (b) Spectral normal absorptance spectra for film-coupled concave grating, concave grating structure without $SiO_2$ spacer, and film-coupled 1D grating structure with the same geometric values from the FDTD simulation.

Fig. 2   Observation planes for the electromagnetic field distribution inside the film-coupled concave grating: (a) the x-y cross section at the center position of the $SiO_2$ spacer (i.e., z = 0), and (b) the x-z plane at the location of half period (i.e., y = $\Lambda/2$). Electromagnetic field distributions from the FDTD simulation at MP1 in the prescribed (c) x-y plane and (d) x-z plane, as well as at MP3 in the prescribed (e) x-y plane and (f) x-z plane.

Fig. 3   Schematics of (a) an LC circuit model and (b) a simplified form for predicting the magnetic resonance condition of MP1.

Fig. 4   Electromagnetic field distributions in prescribed x-y plane at the center position of the $SiO_2$ spacer (i.e., z = 0) for (a) SPP($\pm 1$,0) and (c) SPP($\pm 1, \pm 1$), as well as in the prescribed x-z plane at the location of half period (i.e., y = $\Lambda/2$) for (b) SPP($\pm 1$, 0) and (d) SPP($\pm 1, \pm 1$).

Fig. 5   Spectral normal absorptance for the film-coupled concave grating metamaterial with varied (a) ridge width and (b) grating period.

Fig. 6   Comparison on the resonance conditions between FDTD simulation and LC model for MP1 or dispersion relation for SPP($\pm 1$,0) as a function of (a) ridge width or (b) grating period.

Fig. 7   Contour plots of spectral-directional absorptance of film-coupled concave gratings as a function of frequency and x-direction wavevector for (a) TM waves and (b) TE waves.

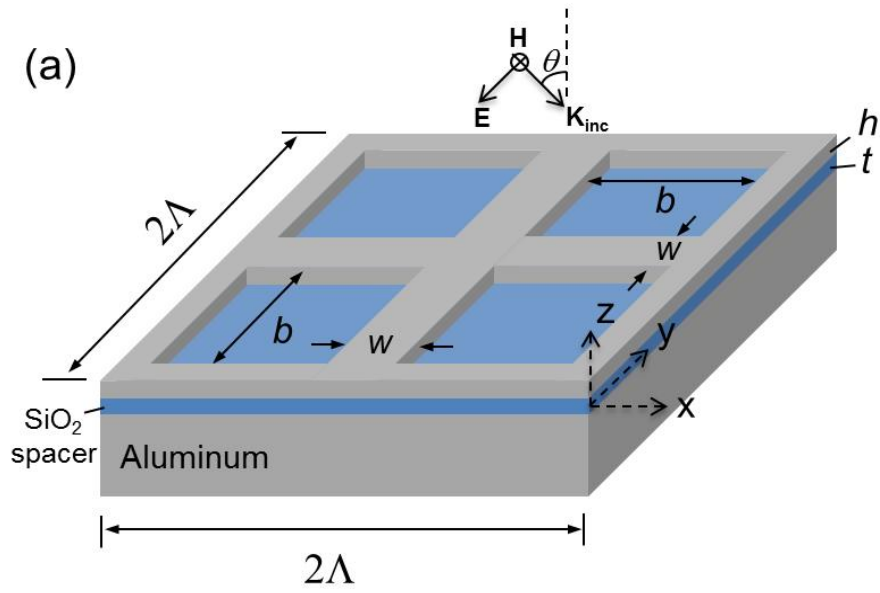

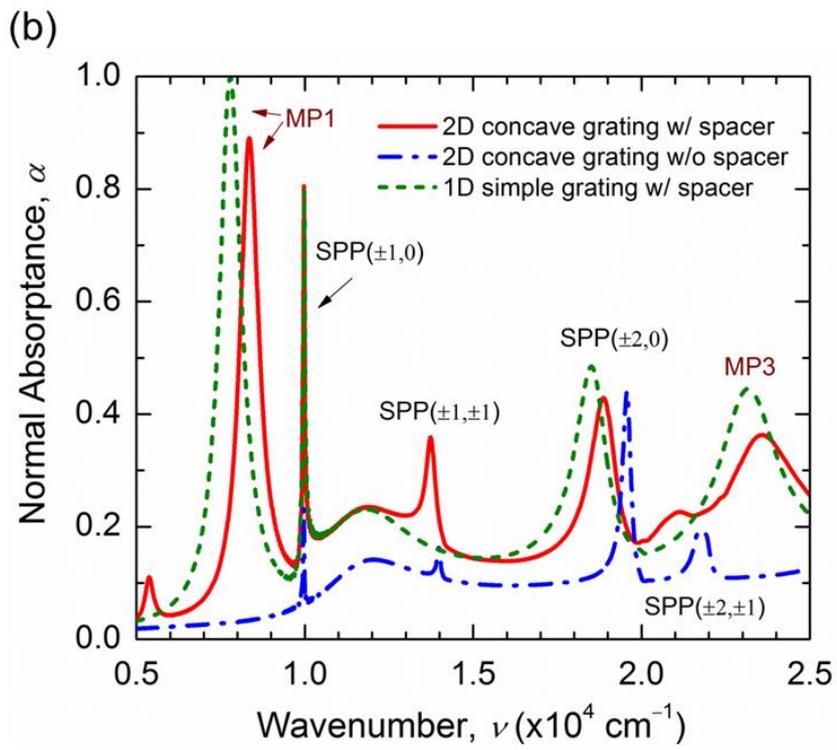

Wang and Wang, FIG. 1

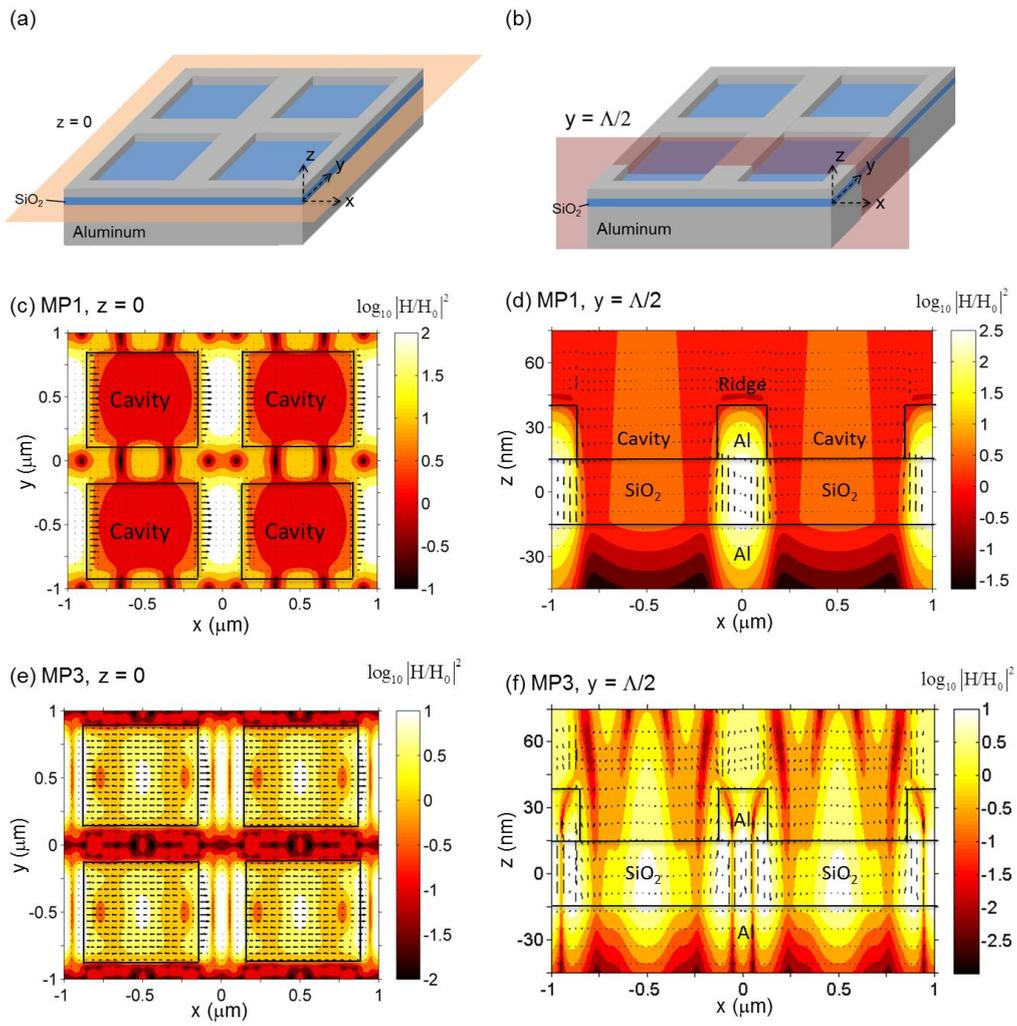

(a) Schematic for LC model

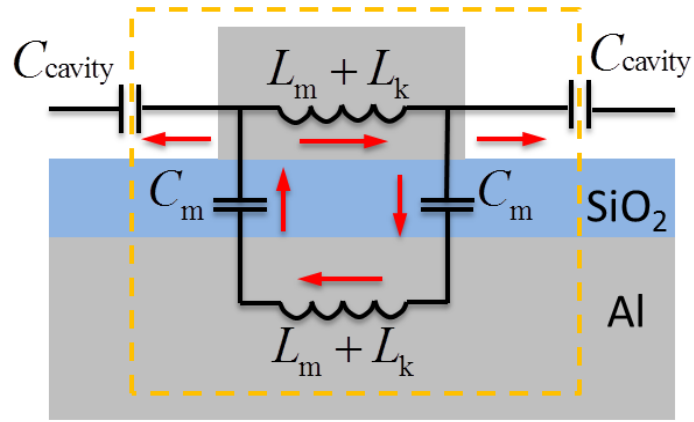

(b) Simplified LC model

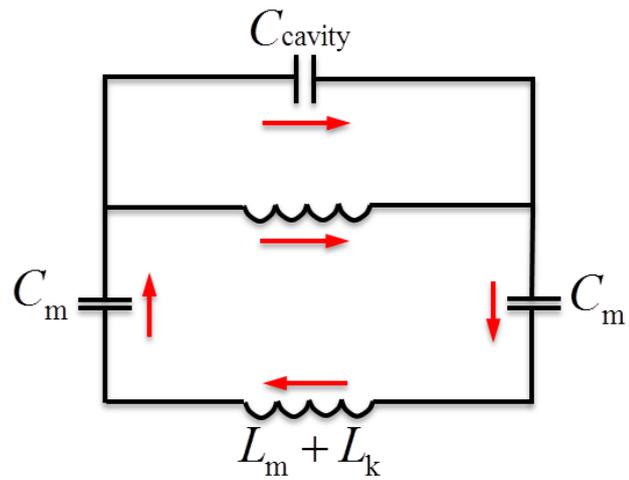

Wang and Wang, FIG. 3

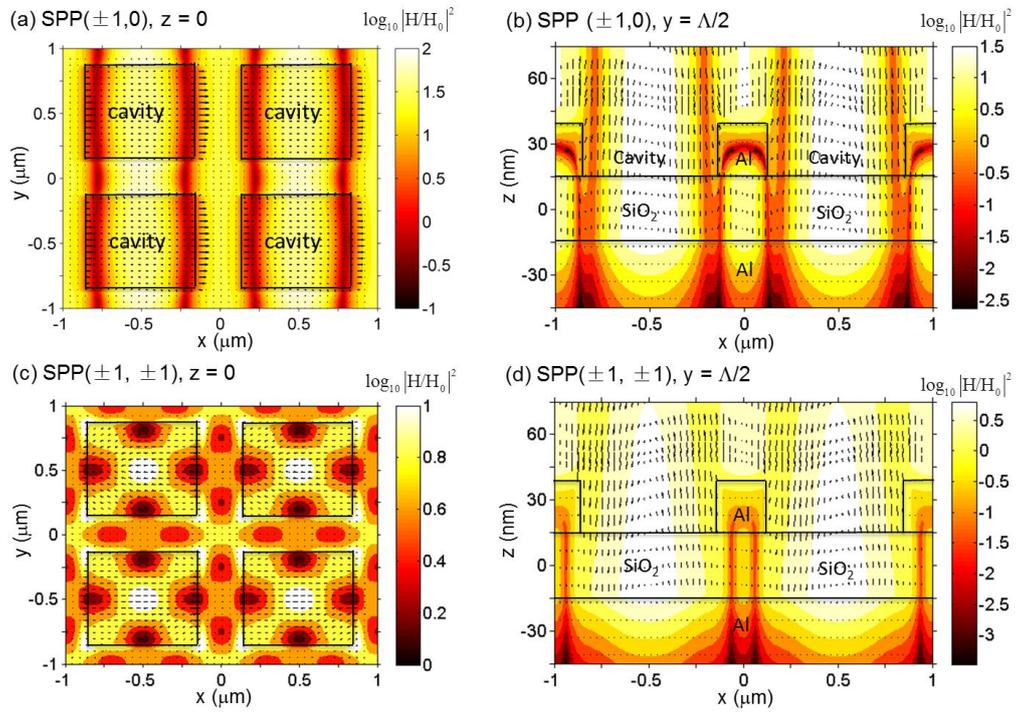

Wang and Wang, FIG. 4

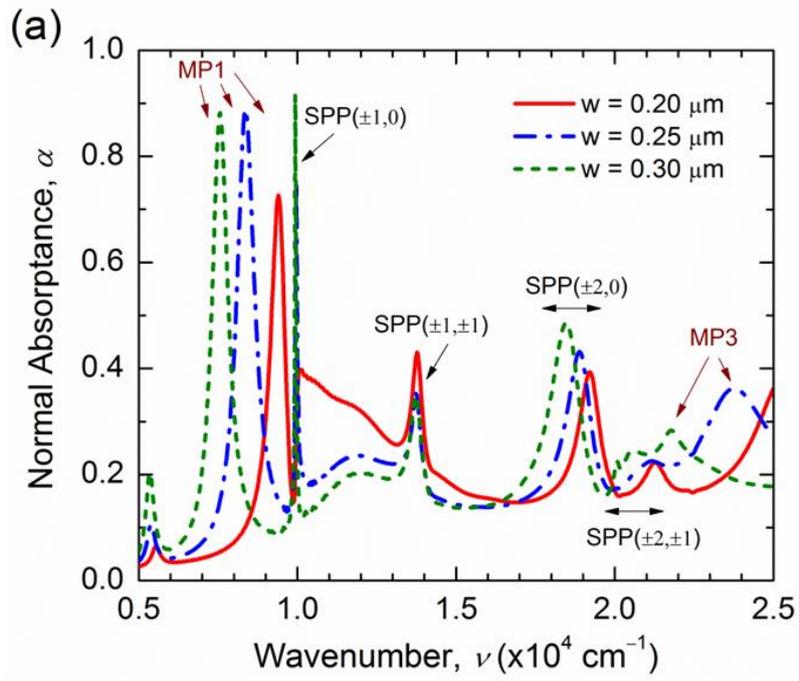

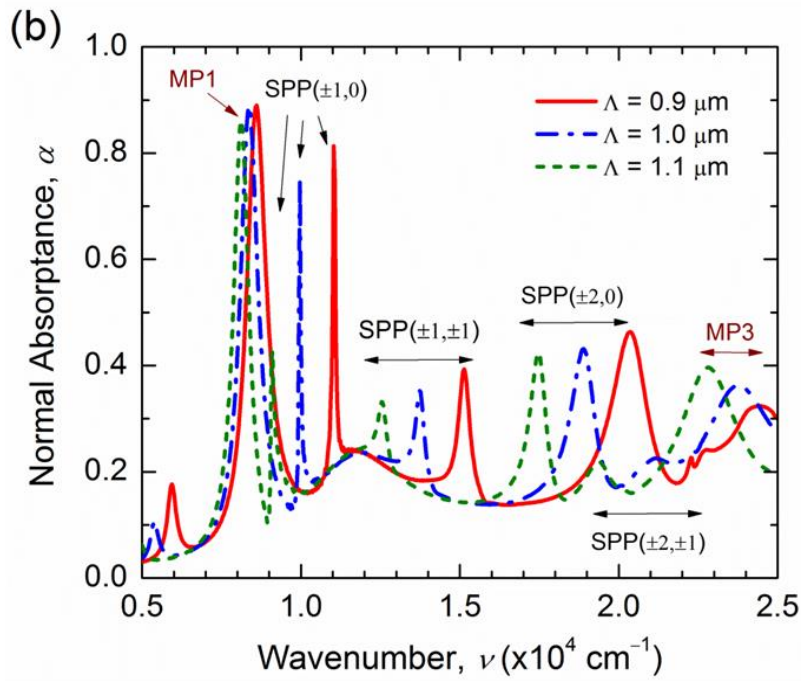

Wang and Wang, FIG. 5

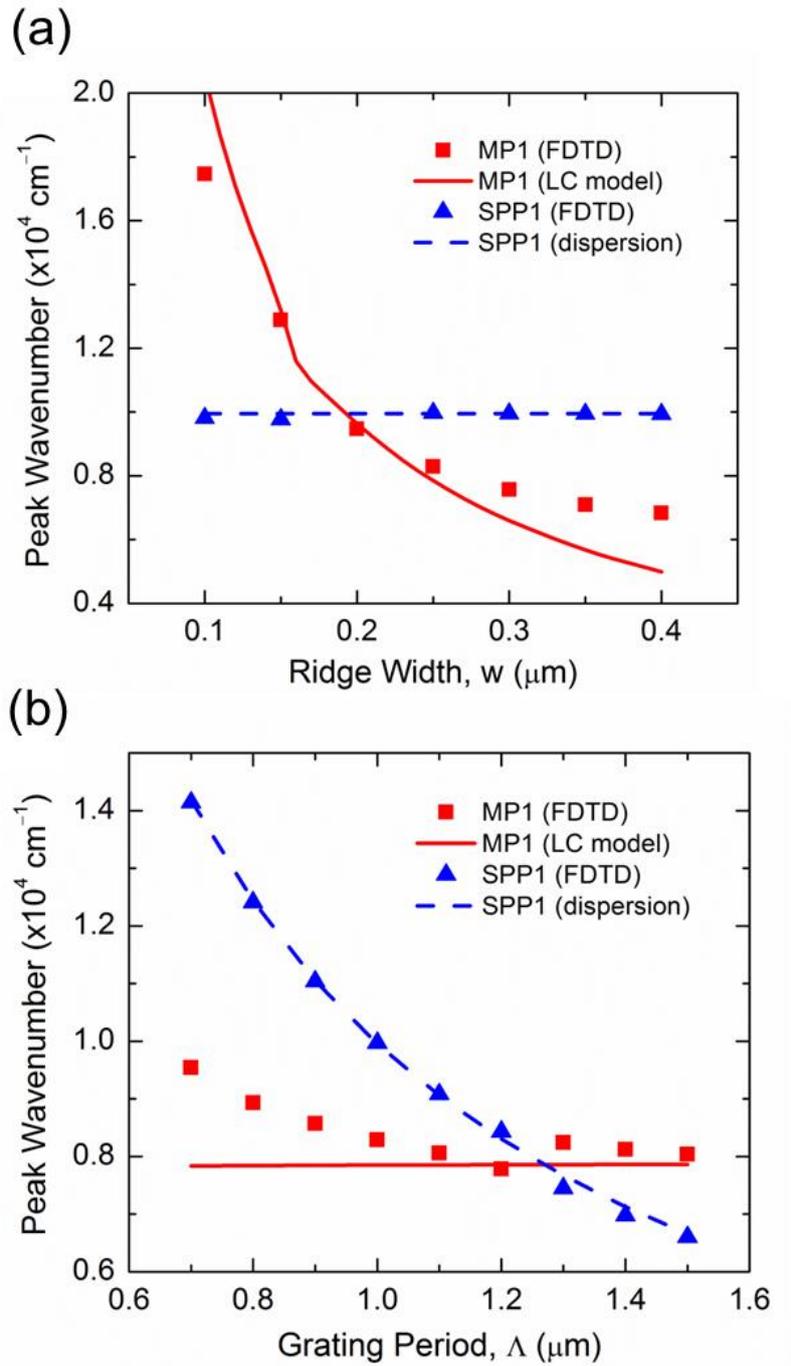

Wang and Wang, FIG. 6

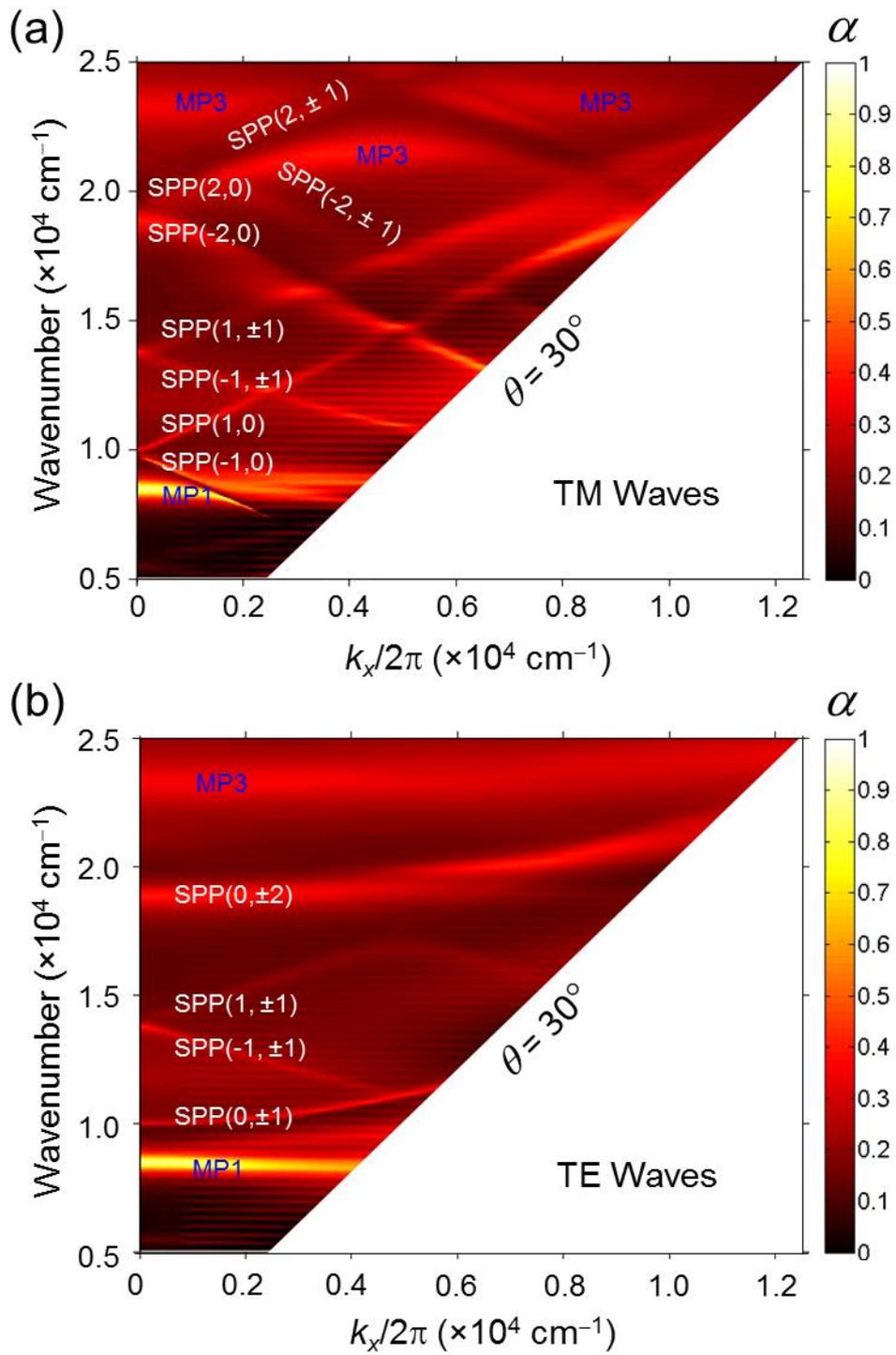